# Effect of Reducing Atmosphere on the Magnetism of $Zn_{1-x}Co_xO$ ($0 \leq x \leq 0.10$) Nanoparticles

M. Naeem, S. K. Hasanain*
*Department of Physics, Quaid-i-Azam University, Islamabad, Pakistan*

M. Kobayashi, Y. Ishida, A. Fujimori
*Department of Physics and Department of Complexity Science and Engineering, University of Tokyo, Kashiwa, Chiba 277-8561, Japan.*

Scott Buzby, S. Ismat Shah
*Department of Physics and Astronomy, University of Delaware, Newark, Delaware 19716, USA*

**Abstract:**

We report the crystal structure and magnetic properties of $Zn_{1-x}Co_xO$ ($0 \leq x \leq 0.10$) nanoparticles synthesized by heating metal acetates in organic solvent. The nanoparticles were crystallized in wurtzite ZnO structure after annealing in air and in a forming gas (Ar95%+H5%). The X-ray diffraction and X-ray photoemission spectroscopy (XPS) data for different Co content show clear evidence for the $Co^{+2}$ ions in tetrahedral symmetry, indicating the substitution of $Co^{+2}$ in ZnO lattice. However samples with x=0.08 and higher cobalt content also indicate the presence of Co metal clusters. Only those samples annealed in the reducing atmosphere of the forming gas, and that showed the presence of oxygen vacancies, exhibited ferromagnetism at room temperature. The air annealed samples remained non-magnetic down to 77K. The essential ingredient in achieving room temperature ferromagnetism in these $Zn_{1-x}Co_xO$ nanoparticles was found to be the presence of additional carriers generated by the presence of the oxygen vacancies.

*corresponding author: skhasanain@qau.edu.pk

# 1. Introduction:

The discovery of diluted magnetic semiconductors (DMS), which are nonmagnetic semiconductors doped with a small amount of some magnetic impurity, has raised tremendous interest in the development of these materials for future technological applications [1]. Due to its wide band gap (3.37 eV) and large excitation energy, transition-metal-doped ZnO has been investigated as a promising DMS for implementing spintronics device concepts [2]. One of the key questions is whether the resulting materials is indeed an alloy of $Zn_{1-x}TM_xO$ (TM=transition metal) or if on the other hand it remains as ZnO with clusters, precipitates, or with second phases that are responsible for the observed magnetic properties [3]. Since the appearance of the paper by Dietl et al [4] predicting the existence of high temperature ferromagnetism (FM) in some magnetically doped wide band gap p-type semiconductors much attention has been focused on these materials. Particularly ZnO and $TiO_2$ doped with different transition metals (Co, Mn, Fe, Ni, Cr, etc) have been the subject of intense research. Most of the theoretical models proposed so far assume TM ions as magnetic impurities and consider only p-type carriers, even though not all the compound semiconductors can be easily doped (without additional elements) with p-type dopants. While several groups have reported the synthesis of p-type ZnO by co-doping method [5, 6], it is still difficult to dope ZnO with p-type dopants at high concentrations. Therefore the synthesis of ZnO with p-type TM doping is not convenient from a practical point of view. However n-type conductivity is in general more advantageous in potential technological application [7]. Recent observations of large anomalous Hall effect in n-type magnetic semiconductors such as Co-doped FeSi [8] and $TiO_2$ [9] suggest that ferromagnetic behavior is also possible in n-type magnetic semiconductors. Thus it is important to search for suitable compositions of n-type ZnO based DMS their preparation routes and the consequent electronic properties. The native defects in ZnO such as oxygen vacancies ($V_o$), zinc vacancies ($V_{Zn}$) and oxygen ($O_i$) and zinc interstitials ($Zn_i$) are understood to be fundamental sources of electron doping [10]. For example the presence of $V_o$ generates free carriers (electrons) that may help in mediating the exchange interaction effects between magnetic impurities. Furthermore, since the spintronic applications envisaged for DMS materials will require miniaturization [11], it is very important to understand the control and variation of the magnetic properties in these systems at the nanoscale. In this study we have investigated Co doped ZnO based DMS nanoparticles for room temperature ferromagnetism. The observation of ferromagnetism at room temperature is examined in the context of the substitution of Co in the lattice and the effects of the annealing environment. We present evidence from X-ray diffraction and X-ray photoemission spectroscopy (XPS) for the presence of Co in the substituted state up to a certain maximum concentration and for the crucial dependence of the observed ferromagnetism on the presence of $V_o$ generated by annealing in a reducing atmosphere.

## 2. Experimental details:

Nanoparticles of $Zn_{1-x}Co_xO$ (x=0 to x=0.10) were synthesized by heating metal acetates in organic solvent following the reported procedure [12]. Zinc-acetate dihydrate $Zn(CH_3COO)_2.2H_2O$ and Co-acetate tetrahydrate $Co(CH_3COO)_2.2H_2O$ were dispersed in a specific volume of ethylenglycol (200ml), and the overall metal concentration was controlled at 0.1M. The suspension was stirred for about 30min and heated at 200ºC for 3 hrs. A blue-green suspension of the oxide was observed for temperatures (*T's*) above 180ºC. After the precipitation of the oxide, the mixture was cooled down to room temperature. The Solid phase was recovered by centrifugation, washed repeatedly with ethanol and finally dried in air at 70ºC. For each concentration of Co, samples were annealed in air and the forming gas (Ar95%+H5%). The sample contained in a quartz boat, was placed inside the tubular furnace and forming gas of purity 99.9% was passed over the samples at 600ºC. The structural characterizations were performed by powder X-ray diffraction (XRD) using CuKα radiation λ=1.5405Å. The samples were also characterized by transmission electron microscopy (TEM) and x-ray photoemission spectroscopy (XPS). XPS measurements were performed using a Gammadata Scienta SES-100 hemispherical analyzer and an Al-Kα (*h*ν=1486.6eV) x-ray source in a vacuum below $1.0 \times 10^{-9}$ Torr at room temperature. Pelletized samples were used for the XPS measurement and clean surfaces were obtained by scraped *in situ* with a diamond file. A vibrating sample magnetometer (VSM) was used to investigate the magnetic properties.

## 3. Results and Discussion:

XRD data of as prepared samples (data not shown) did not show any Bragg peaks indicating that all as prepared samples were amorphous. However when the samples were annealed at 600 ºC in air or in the forming gas, dominant features of a single phase with a hexagonal wurtzite structure were observed as shown in figure 1a. No unidentified peaks were seen in the data. The average particles size for all compositions is about 22nm, as determined from X-ray line broadening using Scherrer formula $D_{hkl} = K\lambda / \beta_{hkl} Cos\theta$, where $D_{hkl}$ is the particles diameter in angstroms, $K$ is the Scherer coefficient equal to 0.89, $\beta$ is the FWHM, and $\lambda$ is the wavelength of X-rays. From these measurements no trace of cobalt metal or cobalt oxides were detectable in any of our doped samples. The position of the XRD peak, corresponding to the 002 plane is seen to be shifted towards lower 2θ values with increasing Co content up to x=0.06. The d(002) values from this peak are plotted as a function of Co content in the inset of figure 1. The lattice d(002) spacing is observed to increase linearly with increasing Co content upto x=0.06 consistent with other reported results on this systems [13,14]. A linear increase of the lattice spacing thus indicates, in accordance with Vegard's law, that at least up to x=0.06, Co ions are substituted in ZnO without changing the Wurtzite structure. On further doping of Co (at x ≥ 0.08) there is a clear decrease of the lattice constant with increasing Co content suggestive of the presence of Co metal clusters. (These may be in addition to the Co ions present in a substitutional role in the ZnO lattice). Bright field transmission electron micrograph (TEM), figure 2, shows that the samples consist generally of spherical particles around 20-50nm in diameter.

XPS studies were conducted to confirm the possible oxidation states of Co ion in the ferromagnetic $Zn_{1-x}Co_xO$ wurtzite structure of the nanoparticles. Figure 3 shows high resolution Co 2p XPS spectra of $Zn_{1-x}Co_xO$, for x = 0.06, 0.08, 0.10. Due to weak signal intensities, it was difficult to resolve the peaks for low Co composition samples. All of these samples show four peaks; the $2p_{3/2}$ and $2p_{1/2}$ doublet and the shake up resonance transitions (satellite) of these two peaks at higher binding energies. The Co $2p_{3/2}$ and $2p_{1/2}$ binding energies are obtained as 781.5eV and 797eV, respectively. These binding energies are very close to those reported for $Co^{+2}$ ions in Co — O [15]. In our case the difference of binding energies between the $2p_{3/2}$ and $2p_{1/2}$ levels is determined to be 15.5±0.1eV, which corresponds well with the value for $Co^{+2}$ homogeneously surrounded by oxygen tetrahedra [16,17]. If Co had existed largely in the form of metal clusters in these samples, the energy difference of these peaks would have been 15.05eV [17]. Therefore, while the presence of Co clusters cannot be ruled out entirely, it appears that at least up till x=0.06 such clustering is at best too small to be within the XPS resolution. However, for samples with x ≥ 0.08, a shoulder starts to appear on the lower energy side of the Co $2p_{3/2}$ peak. This shoulder appears at a binding energy ~778eV below the Co $2p_{3/2}$ peak in figure 3. Generally, core-level binding energy increases with increasing positive valence of ion [18]. It is considered that the Co 2p peak appearing on the lower binding energy side indicates the presence of Co metal cluster. Our data for x≥0.08 therefore suggest that alongwith the $Co^{+2}$ ions in the Zn substituted positions (Co-O), there are also some Co in the form of metallic clusters.

Due to the pronounced difference between the magnetic response of the samples annealed in air and in a forming gas, respectively (discussed later), we investigated the presence of oxygen vacancies in the XPS spectra. In figure 4 we present the O1s XPS spectra of $Zn_{0.94}Co_{0.06}O$. This is the composition with the largest Co content exhibiting no signatures of clustering, both in XRD and XPS. The O1s XPS spectra of both air annealed (figure 4a) and forming gas annealed samples (figure 4b) show a slightly asymmetric peak very close to 530eV. This profile can be fit by two symmetrical peaks, which are normally assigned as low binding energy component (LBEC) and high binding energy component (HBEC) and it has been shown [19] that the HBEC peak develops with increasing loss of oxygen such as by heating in vacuum or by $Ar^+$ bombardment. The development of the HBEC peak obviously leads to the asymmetry of the main peak (LBEC). Our O1s x-ray photoemission spectra yields one clear observation viz. the asymmetry of the LBEC is found to be more pronounced in the samples annealed in forming gas as compared to the samples annealed in air. Furthermore, the relative area under the curve (area of HBEC peak/area of LBEC) is determined to be 0.281±0.002 while for the air annealed samples it is equal to 0.191±0.001. The relatively large contribution of the HBEC peak for the case of annealing in forming gas strongly suggests the presence of more oxygen deficiencies in this case. This is most probably due to the desorption of loosely bound oxygen during annealing in the reducing atmosphere of the forming gas. These oxygen vacancies are expected to generate free carriers (electrons) that can help in mediating the exchange interaction effects between the magnetic impurities.

We now turn to the magnetic properties of $Zn_{1-x}Co_xO$ nanoparticles. The samples annealed in air exhibited no ferromagnetic component down to 77K, the lowest temperature investigated. Interestingly, however, after annealing in a forming gas at 600ºC all samples acquired room temperature ferromagnetism without any observable change in the crystal structure, as revealed by X-ray diffraction, or in the electronic structure observed by the XPS. This is unlike the case for Co doped $TiO_2$ where a reducing atmosphere helps stabilizing the anatase phase that leads to ferromagnetism. The field dependent magnetization (M-H) curves, at room temperature, of all five compositions are shown in figure 5. All samples exhibit RT-FM with the moment (emu/g) monotonically increasing with increasing concentration of cobalt. The maximum saturation magnetization ($M$s) was found to be 3.9emu/g (10% Co doped) among all composition, while the maximum moment per Co atoms was $\mu \sim 0.25\mu_B$/Co (4% Co doped) at room temperature. The values range between 0.141(x=0.02) and 0.533$\mu_B$/Co (x=0.10). These values are comparable with those typically reported in the literature [22]. It is noticeable however that the value of $M$s in our case was quite small (max. 0.533$\mu_B$/Co) in comparison with the value $\mu=3\mu_B$/Co expected from the moment of ionic $Co^{+2}$ in a tetrahedral crystal field. However it is usual in DMS not to recover full magnetization possibly due to antiferromagnetic couplings between some of the neighboring Co ions [21].

A large jump in the magnetic moment is observed at x = 0.10 shown in inset of figure 5. The large moment at x=0.10 may be related, as evidenced by the XPS spectra, to the formation of Co clusters with strong ferromagnetic alignment. In the x=0.08 composition (which appears to be a borderline composition on the basis of magnetization and structural studies) we could also have the possibility of coexistence of divalent Co ions and Co metal clusters as observed in Co 2p XPS spectra. However, in the composition range x ≤ 0.06, where Co was seen to be substituted in the ZnO matrix, the observed ferromagnetic component appears to be arising from Co ions randomly substituted for Zn in the ZnO matrix. The FM correlations in this composition range, where Co ions may be regarded being generally randomly distributed, appear to be connected with the excess electrons arising from the oxygen vacancies [10]. Theoretically, it has been observed that the carrier-mediated ferromagnetism in n-type oxide semiconductors is strongly related to the presence of oxygen vacancies [22]. Whether this indirect exchange interaction is of the RKKY type or due to some other mechanism e.g. virtual magnetic levels [20] close to the band edge, is not understood at the moment.

The coercive field ($H_c$) of 6% Co doped samples was found to be 320Oe at room temperature as shown in figure 4. In thin film sample of similar compositions [16], the coercivity values are much smaller being typically a few tens of Oersteds. The relatively large values we observe are very definitely due to the single domain nature of these nanoparticles and the associated magnetization rotation magnetic mechanism [23]. In nanoparticles, the formation of domain walls is energetically unfavorable below a certain size, depending on the materials, and the particle stays in a single domain configuration. In a single domain particle magnetization reversal can occur only by magnetization rotation (as opposed to domain wall motion) requiring large reversing fields.

Figure 6 shows zero field cooled (ZFC) and field cooled (FC) magnetization data as a function of temperature for the x=0.06 sample at different fields. In the field cooled case the field is applied at room temperature and the data are taken on the way down while in the ZFC case the field is applied at low temperatures where the particle moments may be already blocked along the respective anisotropy axes. The FC and ZFC curves are seen to diverge substantially at low temperatures with the FC data being quite temperature independent for the most part while the zero field data shows a broad maximum. This maximum progressively shifts to a lower temperature with increasing magnetic field (*H*). This behavior is consistent with that expected for magnetic nanoparticles [24] where low temperature leads to spin blocking along the respective anisotropy axes of the particles while field cooling in a sufficient field prevents this blocking. The presence of a large hysteresis even at room temperature indicates that the blocking temperature (and hence the critical temperature) for most of the particles is substantially higher than 300K. Differences between ZFC and FC could in principal also arise due to spin glass type ordering within or between the particles at low temperatures [25]. However the very broad ZFC peak even at the lowest applied field (100Oe) is typical of spin blocking in nanoparticles with a size distribution and unlike that for a typical spin glass.

Two other features are noticeable in the *M(T)* data. Firstly the very large difference between the ZFC values between *H*=100 and 500Oe at room temperature and their coincidence at low temperatures. We understand this large difference at room temperature as occurring due to the value of the applied field exceeding the threshold of 320 Oe (Hc) in this range. For a 100Oe applied field (*H<Hc*) the field is not sufficient to most of the particles while the 500Oe field is quite sufficient. However at low temperatures (say 110K) the coercive field is about 650Oe and both the applied fields (100 and 500Oe) are smaller than the coercivity and the consequent moments are very small and almost the same. Secondly we observe a slight but significant and reproducible upturn in the FC moment at the lowest temperatures, *T*<150K, for the higher fields, 500 and 1000Oe. This appears to be similar to the case of magnetic nanoparticles [26] with a surface layer that undergoes an ordering at low enough temperatures, stabilized by the high magnetic fields. A similar explanation may be valid here as well. Further experiments are underway to identify the origins of this low temperature behavior.

**4. Conclusion**

In conclusion, the structure and magnetic properties of $Zn_{1-x}Co_xO$ (0≤x≤0.10) nanoparticles synthesized by heating metal acetate in organic solvent were studied. No secondary phase was observed in XRD pattern in the whole range of compositions (0≤x≤0.10). The d(0002) value evolution and XPS spectra revealed that the Co ions in $Zn_{1-x}Co_xO$ for x≤0.06 are in the divalent $Co^{+2}$ states with a tetrahedral symmetry. Hence both the structure and the magnetic properties suggest the strong evidence that the Co ions are substituted in ZnO matrix up to x ≤ 0.06. The stark contrast in the magnetic response of the particles annealed in a forming gas to those annealed in air clearly shows that the generation of additional carriers by the oxygen vacancies plays a crucial role, most likely in mediating the exchange interaction between the Co ions. The magnetic behavior of the nanoparticle assembly is consistent with that for a collection of magnetic nanoparticles undergoing blocking along the anisotropy axes and there is evidence of a low temperature ordering of moments possibly in the outer shell of the particles.

Acknowledgement: The work at Quaid-I-Azam University was supported by a research grant No: 20-80/Acad(R)/03 from the Higher Education Commission of Pakistan.

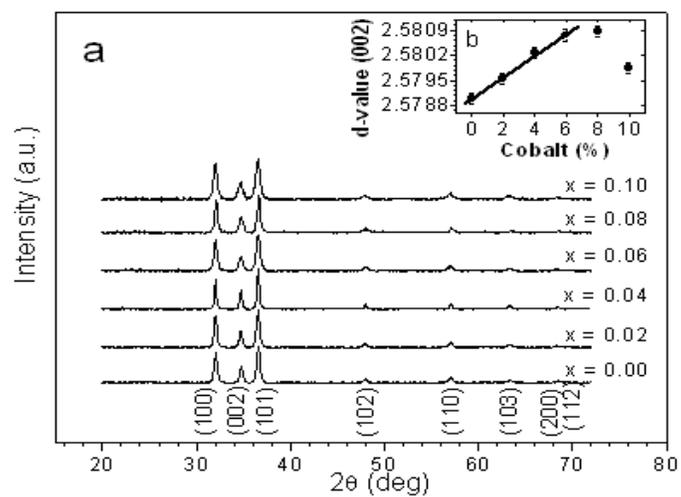

Figure 1 (a) X-ray diffraction patterns of $Zn_{1-x}Co_xO$ (x = 0.00, 0.02, 0.04, 0.06, 0.08, 0.10) nanoparticles. (b) Variation of d(0002) values vs Co concentration.

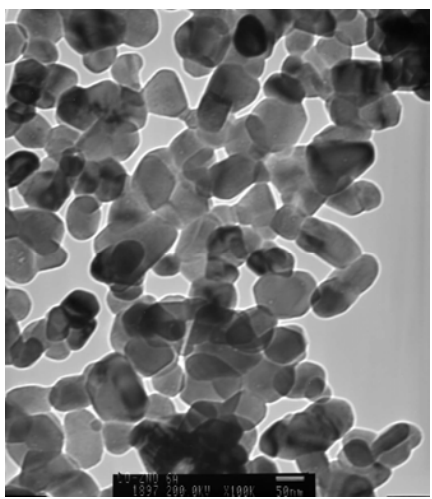

Figure 2 TEM image of $Zn_{0.94}Co_{0.06}O$ nanoparticles.

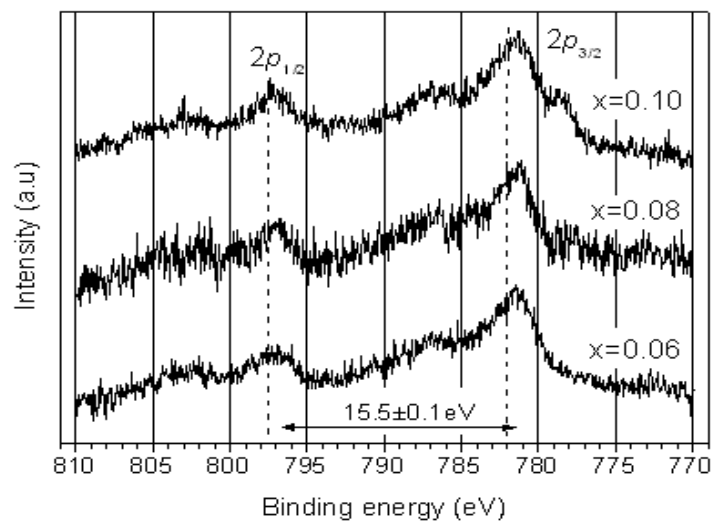

Figure 3 XPS spectra of Co 2*p* core levels as a function of Co concentration (x=0.06, 0.08, 0.10) recorded at room temperature.

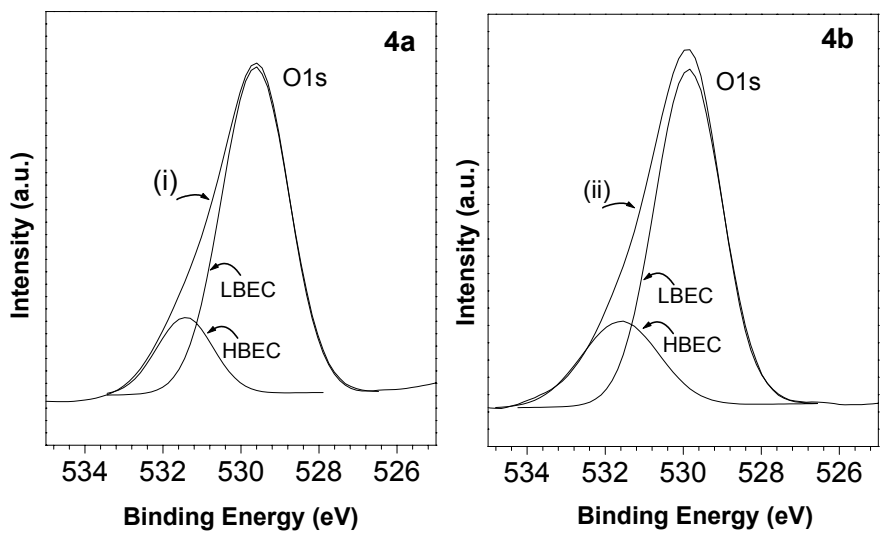

Figure 4 XPS spectra of O1s core level recorded from the samples (a) annealed in air and (b) annealed in forming gas. Fit to the data of curves (i) and (ii) yields the two peaks LBEC and HBEC for each case (see text for details).

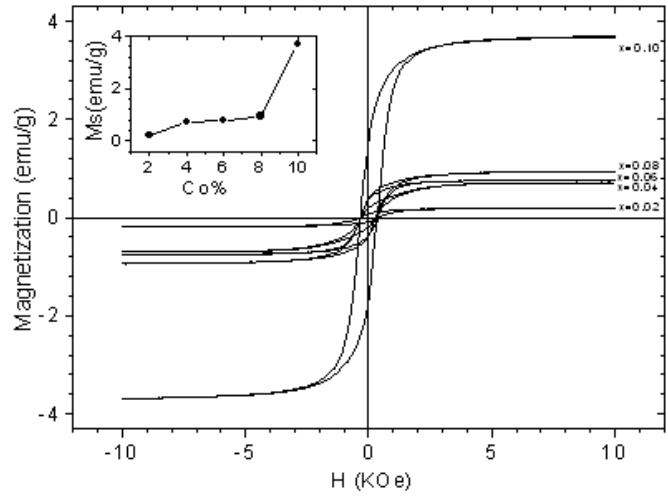

Figure 5 The M-H curve of Co-doped ZnO nanoparticles measured at 300K for different Co concentrations. Inset: Co content (x) dependency of Ms (emu/g).

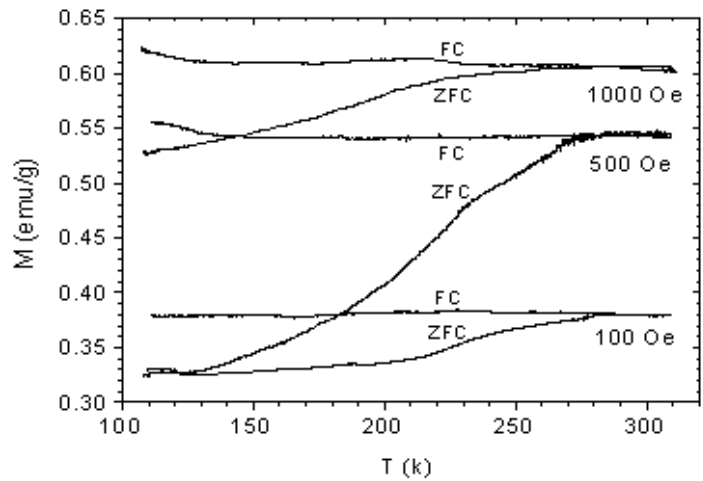

Figure 6 *M-T* curves of $Zn_{0.94}Co_{0.06}O$ nanoparticles at three different magnetic fields.